\documentclass[12pt,a4paper,oneside,showpacs]{revtex4}

\evensidemargin=\oddsidemargin

\begin{document}

\title{Measurement of high order current correlators.}

\author{K.V.\ Bayandin$^{\, a}$, A.V.\ Lebedev$^{\, a,b}$,
G. B. Lesovik$^{\, a}$}

\affiliation{$^{a}$ L.D.\ Landau Institute for Theoretical Physics
RAS, Moscow, 119334, Kosygina str. 2} \affiliation{ $^b$
Theoretische Physik, ETH Zurich, CH-8093 Z\"urich, Switzerland}

\begin{abstract}
      The feasibility of measuring high-order current correlators
      by means of a linear detector is analyzed. Two different types
      of measurements are considered: measurement of fluctuation
      power spectrum and measurement of unequal-time current correlators
      at fixed points in time. In both cases, formally exact expressions
      in terms of Keldysh time-ordered electron current operators are
      derived for the detector output. An explicit time ordering is found
      for the current correlators under the expectation operator used
      in measurements of high order unequal-time current correlators.
      The situation when a detector measures current correlators at
      different points of a conductor is considered.
\end{abstract}

\pacs{03.65.Ta, 05.40.-a, 05.60.Gg, 73.23.-b}

\maketitle

\newpage

\section{Introduction}

Study of noise has recently been gaining popularity. This interest
was quintessentially expressed by R. Landauer in the title of his
paper {\it The Noise Is the Signal}\cite{landauer_98}. It turned
out that measurement of noise (electron current fluctuations) can
be used to determine characteristics of electron transport that
cannot be obtained by measuring only the average current or the
conductance of the system.

The phenomenon of shot noise was predicted by W. Schottky as early
as in the beginning of the past century~\cite{schottky}. Schottky
showed that the variance of the fluctuating current carried by
statistically independent discrete charges is proportional to the
product of the carrier charge with the average current. This fact
can be used to measure the carrier charge in various systems.
Originally, shot noise was measured in electron vacuum
tubes~\cite{pierce}. Much later, measurements of shot noise were
used to determine the quasiparticle charge in edge states for
quantum Hall systems~\cite{glattli,reznikov_97} and the double
electron charge for Cooper pairs in hybrid NS
systems~\cite{kozhevnikov_00}. Furthermore, shot noise measurement
can be used to study the electron correlations due to fermionic
statistics and electron-electron interaction effects (e.g. see
rewiev~\cite{blanter}).

Considerable interest in study of current correlators was also
motivated by prospects of preparation of bipartite entangled states
in superconductors~\cite{les_01} and normal electronic
conductors~\cite{leb}, as well as of entanglement in quantum Hall
effect~\cite{ben}. The corresponding bipartite Bell inequality
(which characterizes the degree of entanglement) is formulated in
terms of second-order current correlators. Moreover, measurement
itself can be a source of entangled electron pairs~\cite{leb}.
In~\cite{ben_04,sim}, a scheme was proposed for preparing an
arbitrary $n$-electron entangled state, which requires measurement
of $n$th-order current correlators.

Furthermore, measurement of second and higher order current
correlators can provide as much information as possible about the
state of a conductor under constraints imposed by quantum theory.
This idea can be implemented in study of the full counting
statistics of transmitted charge. Quantities of this kind have
long since been analyzed in quantum optics in studies of the
counting statistics of photons emitted by various sources (e.g.
see~\cite{mandel}). However, a similar idea regarding electron
systems was proposed only in relatively recent
studies~\cite{LL1,LL2}, where the distribution function was found
for the charge transmitted through a quantum point contact over a
finite time interval.

The past decade has seen a rapid increase in the number of
theoretical studies focused on high-order current correlators, in
particular in diffusive~\cite{lee,nagaev},
chaotic~\cite{blanter_01}, and
interacting~\cite{kindermann_03,golubev_03,golub} electron
systems. Considerable attention was also given to the effects of
temperature~\cite{gutman} and electromagnetic
environment~\cite{nazarov_03} on the third-order current
correlator. A general approach to full counting statistics of
current fluctuations, based on the path integral method, was
recently developed in~\cite{kin,golubev_03}.

Despite numerous theoretical studies, only zero- or finite-frequency
shot noise has so far been amenable to direct measurement
(see~\cite{glattli,reznikov_97,reznikov_95,
schoelkopf_97,kozhevnikov_00}).  In~\cite{hbt}, electron
antibunching was studied in a fermionic Hanbury-Brown-Twiss-type
experiment by measuring electron current cross correlations in a
multiterminal conductor. Only in recent experimental studies, the
third cumulant of electron current fluctuations was measured at low
frequencies in a tunnel junction~\cite{exp31,exp32} and in a quantum
dot in the Coulomb blockade regime~\cite{exp33}.

These measurements of the third cumulant stimulated theoretical
studies where various quantum detectors of current fluctuations
were proposed. They can be divided in two types: (1) detectors
where transitions between discrete energy levels are induced by
current fluctuations~\cite{aguado_00, heikkila,
ojanen,fazio,hassler}; and (2) threshold detectors where
transition from a metastable state is caused by interaction with
current fluctuations~\cite{tobiska,ankerhold}.

Since current operators taken at different times do not commute,
practical measurement of current noise leads to an additional
question about which particular correlation function is measured in
an actual experiment. For second-order current correlators, the
standard prescription is to use the symmetrized correlator $\langle
\hat I(t_2) \hat I(t_1) + \hat I(t_1) \hat I(t_2)
\rangle$~\cite{landau}. Indeed, as shown in~\cite{lesovik2} the
symmetrized correlators at different times can be observed. Note,
that at finite detector temperature, the anti-symmetrized correlator
$i\langle \hat I(t_2) \hat I(t_1) - \hat I(t_1) \hat I(t_2) \rangle$
also contributes to the detector output due to the back action of
the detector on the measured current. An analysis of measurement of
noise power spectrum by means of a resonant LC circuit coupled to
the conductor was presented in~\cite{loosen} (see
also~\cite{imry,aguado_00}). In contrast to the measurement of the
time resolved correlators, it was found that the ground-state
(passive) detector can measure only the positive-frequency noise
power spectrum
$$S(\omega) = \int \langle \hat I(t)\hat I(0) \rangle e^{-i\omega t}
dt,$$ where the current correlator is not time ordered. If the
detector is in an excited state, then the negative-frequency power
spectrum also contributes to the result. This behavior is explained
by the fact that the ground state detector can only absorb energy,
whereas an excited-state detector can transfer energy to the
conductor as well.

An analogous time-ordering problem for current operators under the
expectation operator arises with regard to measurements of
high-order current correlators. In the analyses of full current
statistics presented in~\cite{LL1,LL2}, it was shown that the
transmitted charge cumulants $\langle\langle \hat
Q^n(t)\rangle\rangle$ with $n\geq 3$ depend on the current
measurement method. In particular, when $\hat Q(t)$ is defined as
the integral $\hat Q(t) = \int_0^t \hat I(t^\prime) dt^\prime$
without time ordering, as in the “naive” approach developed
in~\cite{LL1}, the resulting statistics corresponds to a fractional
value of transmitted charge. However, an analysis of a more
realistic measurement scheme using an auxiliary spin-$1/2$
system~\cite{LL2}, revealed that the charge counting statistics (far
from the scattering point at $x\gg x_c$, where for energy
independent scattering $x_c$ is of the order of the Fermi wave
length) is described by a binomial distribution corresponding to an
integer value of the charge. In the latter scheme, the cumulants
$\langle\langle \hat Q^n (t) \rangle\rangle$ are expressed in terms
of Keldysh time-ordered electron current operators. This discrepancy
between current statistics raised a question about which correlator
can be measured in an actual experiment.

In this paper, we theoretically analyze the feasibility of
determination of high-order current correlators by measuring both
fluctuation power spectrum and current correlators at fixed points
in time. The present analysis does not address the dynamics of
electrons in the conductor and behavior of current fluctuations per
se. We consider the joint evolution of the conductor and the
detector and find expressions for the detector output in terms of
correlation functions of current operators, which can be used to
determine all measurable correlation functions of current
fluctuations. Since the arbitrary correlation functions of general
form are considered, the results obtained here apply to measurement
of observables of any quantum system coupled to a detector. However,
the analysis below is focused on the current measurement.
Following~\cite{lesovik2,loosen} we consider harmonic oscillator as
a model of the detector coupled to the conductor at one or several
points. In the general case, measurement of this kind corresponds to
current measurement with a detector dynamics are described by a
linear equation of motion.

\section{GENERAL SCHEMES FOR MEASURING
CURRENT FLUCTUATIONS}

In this section, we describe two different schemes, which measure
the power spectrum of current fluctuations at a finite frequency and
current correlators at fixed times, respectively. We tentatively
treat the detector as an arbitrary quantum system coupled to the
conductor, assuming that measurement of its coordinate $\hat x$
provides information about the magnitude of the current carried by
the conductor. In the sections that follow, the role of a detector
is played by a resonant LC circuit or a single-electron ammeter.
Since either device can be modeled by a harmonic oscillator, the
problem technically reduces to analysis of a harmonic oscillator
driven by an external field (e.g. see~\cite{fenver}), which provides
formally exact expressions for the detector output.

\subsection{Continuous Quantum Measurement}

Suppose that the coupling between the detector and the conductor
is adiabatically switched on at an instant $t_0$ in the past. At
$t>t_0$, a measurement is performed on the detector observable
$\hat x$. The outcome of a quantum measurement on $\hat x$ can be
described only statistically in terms of the distribution function
of the value of $x$, or, equivalently, in terms of moments of the
form $\langle \hat x^n \rangle$, $n\in {\cal Z}$.

This procedure is a typical example of continuous quantum
measurement: the detector continuously interacts with the measured
system, and the detector output is read out either at several points
in time with limited accuracy or only once (see review
in~\cite{mensky_98} and references therein). The outcome is a
time-averaged value of an observable, with an averaging kernel
determined by the detector dynamics. In particular, if the detector
is modeled by a harmonic oscillator, then a cumulant $\langle\langle
\hat x^n\rangle\rangle$ of lowest order in the coupling strength is
proportional to the Fourier transform of an $n$th-order unequal-time
current correlator or, equivalently, to the fluctuation power
spectrum at the oscillator frequency (see below).

The generating function of $\hat x$ defined as:
\begin{equation}
      \chi(\lambda) = Tr_\mathrm{det} \Bigl\{ \hat \rho(t)\,
      \exp(i\lambda \hat x(t))\, \Bigr\},
      \label{1prob}
\end{equation}
where $\hat\rho(t)$ is the detector density matrix and the trace is
taken with respect to detector degrees of freedom. Once
$\chi(\lambda)$ is known, one can find $\langle \hat x^n \rangle$ by
differentiating $\chi(\lambda)$ with respect to $\lambda$:
\begin{equation}
      \langle \hat x^n \rangle = (-i)^n
      \lim_{\lambda \rightarrow 0}
      \bigl( \partial_\lambda^n \chi(\lambda)\bigr),
      \label{cum}
\end{equation}

The generating function is determined by the detector density matrix
$\hat \rho$, which is obtained by tracing out the degrees of freedom
of the measured system from the total density matrix $\hat D$: $\hat
\rho(t) = Tr_\mathrm{sys}\{ \hat D(t)\}$. The density matrix $\hat
D(t)$ is determined by the unitary evolution of the entire system
starting from some point in the past. Suppose that the joint density
matrix of the detector and the conductor at the starting point $t_0$
is a direct product of the form $\hat D(t_0) = \hat \rho_{in} \hat
R_{in}$, where the density matrix $\hat R_{in}$ represents the
initial state of the conductor. We introduce a Hamiltonian $\hat
H_\mathrm{sys}$ to describe free dynamics of the conductor and a
Hamiltonian $\hat H_\mathrm{det}=\hat H_0 +\hat H_\mathrm{int}$, to
describe detector dynamics, where $\hat H_0$ is the free detector
Hamiltonian and $\hat H_\mathrm{int}$ is the coupling Hamiltonian.
In the interaction representation with respect to $\hat H_{sys}$,
the total density matrix at $t>t_0$ has the form:
\begin{equation}
      \hat D(t)  = \hat S(t_0,t) \hat
      \rho_{in} \hat R_{in}(t) \hat S^\dagger(t_0,t),
\end{equation}
where the operator $\hat S(t_0,t)$ describes the evolution of the
detector under the Hamiltonian $\hat H_\mathrm{det}$:
\begin{equation}
      \hat S(t_0,t) = {\cal T}\exp\Bigl( -\frac{i}{h} \int_{t_0}^{t} \hat
      H_\mathrm{det}(t^\prime) dt^\prime \Bigr),
\end{equation}
with ${\cal T}$ is the forward time ordering operator.

We change to the $\hat x$ representation for the total density
matrix and the system evolution operator: $$\hat D(x,y,t)=\langle x|
\hat D(t)|y\rangle,$$  $$\hat S(x_0,t_0;x,t) = \langle x| \hat
S(t_0,t) | x_0\rangle,$$ which remains operator acting on the state
of the conductor. Now, the characteristic function $\chi(\lambda)$
can be represented as:
\begin{equation}
      \chi(\lambda) = Tr_\mathrm{sys} \Bigl\{
      \hat R_{in}(t) \int dx \, e^{i\lambda x} \rho_{in}(x_0,y_0)
      \hat S^\dagger(y_0,t_0;x,t)  \hat S(x_0,t_0;x,t)\Bigr\}.
      \label{gf}
\end{equation}

For $\lambda = 0$, the expression on the right-hand side here is
the influence functional~\cite{fenver}, describing the back-action
of the detector on conductor dynamics.

\subsection{Subsequent Quantum Measurements}

Now, consider another feasible measurement scheme, where the
detector state is measured at subsequent time moments
$t_1<t_2<..<t_n$. According to von Neumann’s projection
postulate~\cite{neumann}, an ideal quantum measurement performed on
$\hat x$ at any $t_i$ projects the detector state at $t_i$ onto an
eigenstate of the operator $\hat x$. This measurement is strong in
the sense that it strongly changes the detector wavefunction at
point $t_i$ and therefore affects its further evolution and
measurement outcomes at $t>t_i$.

First, consider measurements performed on $\hat x$ at two subsequent
moments $t_1$ and $t_2$. The resulting correlation function $\langle
x(t_1) x(t_2) \rangle$ can be expressed in terms of the joint
probability $P(x_1,t_1;x_2,t_2)$ that the values of $\hat x$
measured at $t_1$ and $t_2$ are $x_1$ and $x_2$, respectively:
\begin{equation}
      \langle x(t_1)x(t_2) \rangle = \int dx_1 dx_2\, x_1 x_2\,
      P(x_1,t_1;x_2,t_2).
\end{equation}
In turn, the joint probability $P(x_1,t_1;x_2,t_2)$ is given by
the product: $$P(x_1,t_1;x_2,t_2) =
P(x_2,t_2|x_1,t_1)P(x_1,t_1),$$ where $P(x_2,t_2|x_1,t_1)$ is
probability that the value of $\hat x$ is $x_2$ at $t_2$
conditioned on the outcome $x_1$ of the measurement at moment
$t_1$, and $P(x_1,t_1)$ is the probability to measure the outcome
$x_1$ at moment $t_1$.

The last two probabilities can be determined from principles of
quantum mechanics. If $\hat D(t)$ is the joint density matrix of the
detector and the conductor, then $$P(x_1,t_1) = Tr_\mathrm{sys}\{
\langle x_1|\hat D(t_1)|x_1\rangle\}.$$ The conditional probability
can be found by invoking von Neumann’s projection postulate: at time
moment $t_1$, the total density matrix undergoes an instantaneous
change:
\begin{equation}
      \hat D(t_1^{-0}) \rightarrow \hat D(t_1^{+0})=
      \frac{|x_1\rangle \langle x_1|\hat D(t_1^{-0})
      |x_1\rangle \langle x_1|}{Tr_\mathrm{sys} \{
      \langle x_1| \hat D(t_1^{-0}) |x_1\rangle\}},
      \label{reduc}
\end{equation}
where the denominator is introduced to normalize the reduced
density matrix. At $t>t_1$, the reduced matrix is described by a
unitary evolution operator: $$\hat D(t_2) = \hat S(t_1,t_2)\hat
D(t_1^{+0})\hat S^\dagger(t_1,t_2).$$ Thus, the conditional
probability $P(x_2,t_2|x_1,t_1)$ is expressed as follows:
\begin{equation}
      P(x_2,t_2|x_1,t_1) = Tr_\mathrm{sys} \Bigl\{
      \hat S(x_1,t_1;x_2,t_2)\,
      \frac{\hat D(x_1,x_1,t_1)}{Tr_\mathrm{sys}\{
      \hat D(x_1,x_1,t_1)\}}\, \hat
      S^\dagger(x_1,t_1;x_2,t_2)\Bigr\},
\end{equation}
where $\hat D(x,y,t)=\langle x|\hat D(t)|y\rangle$ is the total
density matrix in the basis of eigenfunctions of the observable
$\hat x$.

Taking the product of the conditional probability on $P(x_1,t_1)$,
we find that the normalization factor cancels out. The result is
\begin{equation}
      P(x_1,t_1;x_2,t_2) = Tr_\mathrm{sys} \bigl\{
      \hat S(x_1,t_1;x_2,t_2) \hat D(x_1,x_1,t_1)
      \hat S^\dagger(x_1,t_1;x_2,t_2)\bigr\}.
\end{equation}

Since the correlation function $\langle x(t_1)x(t_2)\rangle$ is
calculated by averaging over all possible outcomes at $t_1$ and
$t_2$, the measurement performed at $t_1$ can be interpreted as
the instantaneous diagonalization of $\hat D(x,y,t)$ in the basis
of the measured observable,
\begin{equation}
      \hat D(x,y,t_1^{-0}) \rightarrow
      \hat D(x,x,t_1^{+0}),
      \label{proj}
\end{equation}
which is an equivalent formulation of von Neumann’s projection
postulate.

In the general case of subsequent measurements, the correlation
function of the detector outputs at times $t_i$ can be calculated
as:
\begin{equation}
      \langle x(t_1) ... x(t_n) \rangle = \int
      \prod\limits_{i=1}^n x_i\,dx_i\,
      P(\{x_i,t_i\}),
\end{equation}
where $P(\{x_i,t_i\})$ is the joint probability that the detector
coordinates at $t_1,...,t_n$ are $x_1,...,x_n$ respectively. For a
single measurement of the detector coordinate $\hat x$, the
probability $P(x_1,t_1)$ is expressed in terms of the diagonal
elements of the detector density matrix at time $t_1$:
$$P(x_1,t_1)=Tr_\mathrm{sys}\{ \hat D(x_1,x_1,t_1)\}.$$ By analogy,
the probability $P(\{x_i,t_i\})$ for subsequent measurements can be
expressed in terms of diagonal elements of the density matrix к
\begin{eqnarray}
      \hat D(\{x_i,x_i,t_i\}) &=& \int dx_0dy_0\,
      \hat S(x_{n-1},t_{n-1};x_n,t_n)\, ...\, \hat
      S(x_0,t_0;x_1,t_1)\, \rho_{in}(x_0,y_0)\hat R_{in}
      \nonumber
      \\
      &\times& \hat
      S^\dagger(y_0,t_0;x_1,t_1)\,...\, \hat
      S^\dagger(x_{n-1},t_{n-1};x_n,t_n),
      \label{D}
\end{eqnarray}
where the values of $\hat x$ at $t=t_i$ have to be equal to each
other under forward and backward time evolution. Thus, the detector
density matrix at the times of measurements, $t=t_i$, is diagonal in
the basis of eigenstates of $\hat x$. Therefore, the required
probability is
\begin{equation}
      P(\{x_i,t_i\}) = Tr_\mathrm{sys}\{ \hat D(\{x_i,x_i,t_i\})\}.
      \label{Pn}
\end{equation}

Note that the correlation function defined in this manner must be
real-valued, because each measurement outcome is a real number by
virtue of von Neumann’s projection postulate. This correlation
function should be distinguished from the correlation function
$\langle \hat x(t_1) ... \hat x(t_n) \rangle$ of coordinate
operators, which may not be real-valued in the general case.

In this approach, each measurement projects the state of the system
onto an eigenstate of the measured observable, as in von Neumann’s
ideal measurement. However, it is not the only type of feasible time
resolved measurements. For example, the theory of photon detection
in quantum optics assumes that the photodetector is “reset” to the
ground state after each count and then switched back to the standby
mode~\cite{mandel}. In distinction to~(\ref{reduc}), this
measurement of $x(t_i)=x_i$ formally corresponds to an instantaneous
transformation of the density matrix bringing the system into the
next state,
\begin{equation}
      \hat D(t_i^-) \rightarrow \hat D(t_i^+)=
      \hat \rho_i \otimes \frac{\langle x_i| \hat
      D(t_i^-)|x_i\rangle}{Tr_\mathrm{sys}\{
      \langle x_i| \hat D(t_i^-)|x_i\rangle\}},
\end{equation}
which corresponds to the direct product of the detector density
matrix $\hat \rho_i$ with the conductor one. The density matrix
$\hat \rho_i$ represents the detector state in standby mode at
time $t_i$. The corresponding correlation function of $\hat x$ is
expressed as:
\begin{eqnarray}
      &&\langle x(t_1) ... x(t_n) \rangle = Tr_\mathrm{det}\{ \hat x\,
      \hat S(t_{n-1},t_n)\hat \rho_{n-1}\times... \times
      Tr_\mathrm{det}\{ \hat x\, \hat S(t_1,t_2)\hat \rho_1
      \nonumber
      \\
      &&\times\,
      Tr_{det} \{ \hat x\, \hat S(t_0,t_1)\hat \rho_{in}\hat R_{in}
      \hat S^\dagger(t_0,t_1)\}\hat S^\dagger(t_1,t_2)\}\times...
      \times \hat  S^\dagger(t_{n-1},t_n) \bigr\}.
      \label{xxw}
\end{eqnarray}

In a sense, this type of measurement is more natural than von
Neumann’s ideal measurement. Indeed, the procedure described here
assumes that the detector itself is a quantum-mechanical system
whose state must also be measured. The state of the detector can be
measured by coupling it to some macroscopic device. The coupling
must be sufficiently strong for the detector state to be determined
over a time interval much shorter than the interval between
subsequent measurements. The resulting detector state may be
different from $|x_i\rangle$, for example, in the ground state. This
occurs when a detector excited to a metastable state by interaction
with the measured system loses energy to the macroscopic device.

\section{MEASUREMENT OF NOISE POWER SPECTRUM}

Consider a measurement of the power spectrum of current
fluctuations using a resonant LC circuit inductively coupled to a
quantum conductor as a detector. Suppose that the observable to be
measured is the electric charge $\hat q(t_1)$ stored in the
capacitor at a point in time $t_1$. As a model of the LC circuit,
consider a weakly damped harmonic oscillator driven by an external
force $\hat J(t)=\sum_i \alpha_i \hat I_i(t)$. The total
Hamiltonian is
\begin{equation}
      \hat H_\mathrm{det} = \frac{L\hat j^2}2 + \frac{\hat q^2}{2C} +
      \hat j \sum_i \alpha_i \hat I_i(t) ,
      \label{hLC}
\end{equation}
where $L$ and $C$ denote the inductance and capacitance of the
circuit, the operators $\hat j$ and $\hat I_i$ represent the
currents through the LC circuit and through the conductor at
coordinate $x_i$; and $\alpha_i$ denotes the mutual inductance
between the LC circuit and the conductor in the neighborhood of
$x_i$. The operators $\hat q$ and $\hat j$ can be expressed in
terms of bosonic creation and annihilation operators:
\begin{equation}
      \hat q = \Bigl( \frac{\hbar}{2\Omega L} \Bigr)^{1/2} \bigl(
      \hat a^\dagger + \hat a \bigr), \quad
      \hat j =i \Bigl( \frac{\hbar \Omega}{2L} \Bigr)^{1/2} \bigl(
      \hat a^\dagger - \hat a \bigr),
\end{equation}
where $\Omega = 1/\sqrt{LC}$ is the resonant frequency of the LC
circuit.

For the harmonic oscillator with Hamiltonian~(\ref{hLC}), the
evolution operator in the $\hat q$ representation can be found in
explicit form:
\begin{eqnarray}
      &&\hat S(q_1,q_2,T) =S_0(q_1,q_2,T)\,
      {\cal T}\exp\Biggl( \frac{iL\Omega}{\hbar\sin\Omega T}
      \Biggl\{ \frac{q_1}{L} \int\limits_{t_1}^{t_2} \hat J(t)
      \cos\Omega(t_2-\!t) dt
      \label{Ke}
      \\
      &&-\frac{q_2}{L} \int\limits_{t_1}^{t_2}
      \hat J(t)  \cos\Omega(t-\!t_1)dt +
      \frac{1}{L^2} \int\limits_{t_1}^{t_2} dt \int\limits_{t_1}^t ds \,
      \hat J(t) \hat J(s) \, \cos\Omega(t_2-\!t) \cos\Omega(s-\!t_1) \Bigr\}
      \Biggr),
      \nonumber
\end{eqnarray}
where $T=t_2-t_1$, and the transition amplitude for a free
harmonic oscillator is expressed as
\begin{equation}
      S_0(q_1,q_2,T) = \Bigl( \frac{L\Omega}{2\pi i \hbar
      \sin\Omega T} \Bigr)^{1/2}
      \exp\Bigl( \frac{iL\Omega}{2\hbar \sin \Omega T}
      \bigl( (q_1^2+q_2^2)\cos\Omega T- 2 q_1 q_2 \bigr) \Bigr).
      \label{S0}
\end{equation}

Assuming that the LC circuit prior to measurement (at $t_0
\rightarrow -\infty$) is in equilibrium at temperature $\Theta$
and and performing the integral in~(\ref{gf}) over the initial and
final detector coordinates, we find that $\chi(\lambda)$ is the
product of the equilibrium characteristic function
$\chi_0(\lambda)$ with the characteristic function of excess
capacitor-charge fluctuations:
\begin{equation}
      \chi(\lambda)= \chi_0(\lambda) \left\langle
      {\cal T}_{\pm} \exp \left[
      \frac{i\lambda}{\hbar} \int_{t_0}^{t_1} \bigl(
      G(t_1-t) \hat J_{\scriptscriptstyle +}(t) +
      G^*(t_1-t) \hat J_{\scriptscriptstyle -}(t)
      \bigr)\, dt \right]
      F[\hat J_{\scriptscriptstyle +},\hat J_{\scriptscriptstyle -}]
      \right\rangle ,
      \label{gfc}
\end{equation}
where the influence functional of the detector $F[\hat
J_{\scriptscriptstyle +},\hat J_{\scriptscriptstyle -}]$ is
expressed as
\begin{equation}
      F[\hat J_{\scriptscriptstyle +},
      \hat J_{\scriptscriptstyle -}] =
      \exp \left[ -\frac1{\hbar^2} \int_{t_0}^{t_1} dt
      \int_{t_0}^t ds\, \bigl( \hat J_{\scriptscriptstyle +}(t) -
      \hat J_{\scriptscriptstyle -}(t)\bigr)
      \bigl(\Gamma(t-s)\hat J_{\scriptscriptstyle +}(s) -
      \Gamma^*(t-s) \hat J_{\scriptscriptstyle -}(s)\bigr)
      \right],
      \label{F}
\end{equation}
the time-ordering operator ${\cal T}_\pm$ is introduced to order
the current operators $\hat J_{\scriptscriptstyle \pm}(t)$ forward
or backwards in time, and averaging is performed over the detector
states. The response functions in~(\ref{gfc}) are defined as
follows:
\begin{eqnarray}
      G(t) &=& \Bigl( \frac{\hbar}{2L} \Bigr) \bigl( (N+1)e^{-i\Omega
      t} - N e^{i\Omega t} \bigr)\\
      \Gamma(t) &=& \Bigl( \frac{\hbar\Omega}{2L} \Bigr)
      \bigl( (N+1)e^{-i\Omega  t} + N e^{i\Omega t} \bigr),
\end{eqnarray}
where $N=1/(e^{\hbar\Omega/\Theta}-1)$ is the bosonic occupation
number for the oscillator mode. The equilibrium characteristic
function $\chi_0(\lambda)$ corresponds to Gaussian charge
fluctuations with variance $\delta q^2=(\hbar/L\Omega)(N+
\frac12)$:
\begin{equation}
      \chi_0(\lambda) =
      \exp \Bigl( -\frac12 \lambda^2 \delta q^2 \Bigr).
\end{equation}

Expressions~(\ref{gfc}) and~(\ref{F}) can be rewritten in more
compact form in terms of the Keldysh Green’s functions
$$G_K(t-t^\prime) = -i\langle {\cal T}_K \, \hat q(t) \hat
j(t^\prime) \rangle,$$ $$\Gamma_K(t-t^\prime) = \langle {\cal T}_K
\, \hat j(t) \hat j(t^\prime) \rangle,$$ where the operator ${\cal
T}_K$ denotes time-ordering along the Keldysh contour consisting
of forward and backward branches $(t_0, t_1)$ and $(t_1,t_0)$:
\begin{eqnarray}
      &&\chi(\lambda) = \chi_0(\lambda)
      \Biggl\langle {\cal T}_{K}
      \exp \left[\frac{i\lambda}{\hbar}
      \int_K G_K(t_1-t) \hat J(t)\,
      dt\right]
      \nonumber
      \\
      &&\qquad \quad \times
      \exp \left[ - \frac1{2\hbar^2}
      \int\int_K \Gamma_{K}(t-s) \hat J(t) \hat
      J(s)\, dt ds
      \right]
      \Biggr\rangle.
      \label{chiq}
\end{eqnarray}

This expression provides a formally exact description of capacitor
charge statistics for the LC circuit. The characteristic function
$\chi(\lambda)$ can hardly be calculated in explicit form, but
expression~(\ref{chiq}) makes it possible to develop a
perturbation theory in the parameter $\alpha/L\ll 1$ by
representing the influence functional $F[ \hat
J_{\scriptscriptstyle +}, \hat J_{\scriptscriptstyle -}]$ as a
Taylor series expansion.

The analysis that follows focuses on the limit case when the
back-action of the detector on the measured system, described by
the influence functional given by~(\ref{F}), is negligible.
Setting $F[\hat J_{\scriptscriptstyle +}, \hat
J_{\scriptscriptstyle -}]=1$, we obtain:
\begin{equation}
      \chi(\lambda)= \chi_0(\lambda) \left\langle {\cal T}_{K}
      \exp \left[\frac{i\lambda}{\hbar}
      \int_K G_{K}(t_1\!-t) \hat J(t)
      dt\right] \right\rangle.
      \label{chiq2}
\end{equation}

First, we examine the autocovariance of the capacitor charge
fluctuations. Using~(\ref{cum}), we obtain:
\begin{equation}
      \langle \hat q^2 \rangle = \delta q^2 +\frac1{\hbar^2}
      \left\langle {\cal T}_K \Bigl( \int_K G_K(t_1-t) \hat J(t)
      dt \Bigr)^2 \right\rangle,
      \label{per2}
\end{equation}
where the first and second terms correspond to equilibrium and
excess fluctuations, respectively.

It should be noted here that perturbation series expansions contain
secular terms~\cite{sek}. Indeed, even though the exact expression
for $\chi(\lambda)$ given by eq.~(\ref{chiq}) formally yields a
finite result for any $\lambda$, each term in the perturbation
series generally diverges as $t_1\rightarrow \infty$. This is
obviously true even for second-order quantities: in the stationary
case, the current-current correlator in~(\ref{per2}) is determined
only by a time difference, and the resulting integral diverges as
one of the remaining time variables goes to infinity.

These divergences can be eliminated by using weakly damped Green’s
functions. In physical terms, this means that a real detector not
only interacts with the conductor, but also loses energy to the
environment.

However, the introduction of an external heat bath is not a
physical necessity. Indeed, its role can be played by the detector
itself. The process can qualitatively be described as follows. An
interaction with the conductor brings the detector into a steady
state that can be strongly overheated as compared to its initial
state. Any further excited state of the detector has a finite
lifetime, because the detector will tend to lose energy to the
conductor.

The degree of detector overheating can be estimated
perturbatively. For the detector starting from the state with
occupation number $N$, we use perturbation theory to calculate the
probabilities of excitation to the state $N+1$ and de-excitation
to the state $N-1$ over a time interval $\Delta t\gg \Omega^{-1}$:
\begin{eqnarray}
      P_{N+1}(\Delta t) &=& \frac{\alpha^2\Omega\Delta t}{2\hbar
      L}\, (N+1)\, S^{(2)}(\Omega),
      \\
      P_{N-1}(\Delta t) &=& \frac{\alpha^2 \Omega \Delta t}{2\hbar
      L}\, N\, S^{(2)}(-\Omega).
\end{eqnarray}
In perturbation theory, these probabilities are low. The power
spectrum of current fluctuations is defined as
\begin{equation}
      S_{ij}^{(2)}(\omega) = \int \langle \hat I_i(t_1) \hat I_j(t_2)
      \rangle \, e^{-i\omega(t_1-t_2)} d(t_1-t_2).
\end{equation}

Using the balance condition $P_{N-1}\geq P_{N+1}$, we estimate the
steady-state occupation number as
\begin{equation}
      \bar N \sim
      \frac{S^{(2)}(\Omega)}{S^{(2)}(-\Omega)-S^{(2)}(\Omega)}.
\end{equation}
This quantity may be large even in the case of weak coupling
between the detector and the conductor, which becomes obvious as
the excitation energy tends to zero ($\Omega \rightarrow 0$).

However, we assume below that the detector is coupled to the
environment, introducing the factor $e^{-\eta|t|}$ into integrals
in all expressions analogous to~(\ref{per2}). To estimate the
leading-order contribution to the excess capacitor-charge variance
$\langle \hat q^2\rangle$, we express the average over detector
states in~(\ref{per2}) in terms of integrals of the noise power
spectrum:
\begin{equation}
      \langle \hat q^2 \rangle = \sum_{ij}
      \frac{\alpha_i \alpha_j}{(2L)^2} \Bigl\{ (N+1) \int
      \frac{d\omega}{\pi} \frac{S_{ij}^{(2)}(\omega)}{(\omega -\Omega)^2
      +\eta^2} - N \int \frac{d\omega}{\pi}
      \frac{S_{ij}^{(2)}(\omega)}{(\omega +\Omega)^2
      +\eta^2} \Bigr\}.
\end{equation}
In the weak-damping limit ($\eta \ll \Omega$), we have
$\eta/(x^2+\eta^2) \rightarrow \pi \delta(x)$, and the result is
expressed in terms of the positive- and negative-frequency power
spectra, $S_{ij}^{(2)}(\Omega)$ and $S_{ij}^{(2)}(-\Omega)$:
\begin{equation}
      \langle \hat q^2\rangle = \sum_{ij} \frac{\alpha_i
      \alpha_j}{(2L)^2}\, \frac1{\eta}\, \Bigl\{ S_{ij}^{(2)}(\Omega) +
      N ( S_{ij}^{(2)}(\Omega)-S_{ij}^{(2)}(-\Omega)) \Bigr\}.
      \label{S2}
\end{equation}
Expression~(\ref{S2}) extends the results presented
in~\cite{loosen} to the case when the detector is coupled to the
conductor at several points. The result for current measurement at
two different points was presented in~\cite{lesovik_05} and
calculated later perturbatively in~\cite{martin_06}.

It is clear that negative-frequency current fluctuations
contribute to this result only at finite temperatures ($N>1$),
whereas positive-frequency fluctuations contribute even at zero
detector temperature. As noted in~\cite{loosen,imry} the reason is
that energy transfer at zero temperature is possible only from the
conductor to the detector, and the intensities of such processes
are proportional to $S_{ij}^{(2)}(\Omega)$. At a finite
temperature, the detector can be in an excited state from the
start and transfer energy to the conductor at a rate proportional
to $S_{ij}^{(2)}(-\Omega)$.

The proposed method can be used to measure any electron current
correlation function. However, fluctuations of capacitor charge
are dominated by fourth-order correlators. Indeed, it can easily
be shown that the lowest order contribution to the mean stored
charge,
\begin{equation}
      \langle \hat q \rangle = \frac1{\eta} \sum_i
      \frac{\alpha_i}{L}\, \Bigl( \frac{\eta}{\Omega} \Bigr)^2
      \,\langle \hat I_i\rangle,
      \label{q2}
\end{equation}
is small since $\eta/\Omega \ll 1$. An analogous result holds true
for any contribution to charge fluctuations containing odd-order
correlators.

Now, consider measurement of fourth-order correlators. Retaining
only the lowest order nonvanishing contribution to the
fourth-order cumulant $\langle \langle \hat q^4 \rangle \rangle =
\langle \hat q^4 \rangle - 3 \langle \hat q^2 \rangle^2$, and
using characteristic function~(\ref{chiq2}), we obtain:
\begin{equation}
      \langle \langle \hat q^4 \rangle \rangle = \frac1{\hbar^4}
      \left\langle {\cal T}_K  \Bigl( \int_K G_K(t_1\!-t) \hat J(t)
      dt \Bigr)^4 \right\rangle -3 \langle\hat q^2\rangle^2.
\end{equation}

For simplicity, suppose that $N=0$ at the starting point. Then,
the dominant contribution to capacitor charge fluctuations
corresponds to the following time ordered average of current
operators:
\begin{equation}
      \langle\langle \hat q^4\rangle\rangle = \frac{6}{(2L)^4}
      \int\limits_{-\infty}^0 ds_1ds_2 dt_1 dt_2 \,
      e^{i\Omega(t_1+t_2-s_1-s_2)}
      \left\langle {\cal T}_-\{
      \hat J(s_1) \hat J(s_2)\} {\cal T}_+ \{ \hat J(t_1) \hat
      J(t_2) \}\right\rangle.
\end{equation}
The fourth-order unequal-time current correlator contained in this
expression can be represented as the sum of an irreducible
correlator and a number of reducible ones: $$\langle \hat I_1 \hat
I_2 \hat I_3 \hat I_4 \rangle = \langle\langle \hat I_1 \hat I_2
\hat I_3 \hat I_4\rangle\rangle + \langle \hat I_1 \hat I_2\rangle
\langle \hat I_3 \hat I_4\rangle + \langle \hat I_1 \hat
I_4\rangle \langle \hat I_2 \hat I_3\rangle + \langle \hat I_1\hat
I_3\rangle \langle \hat I_2 \hat I_4\rangle+...,$$ where all
odd-order correlators are dropped. Since steady-state current
correlators depend only on time differences, the spectral density
of the fourth-order irreducible correlator can be represented as
\begin{equation}
      S_{ijkl}^{(4)}(\omega_1,\omega_2,\omega_3) = \int
      \prod_{i=1}^3  d(t_i\!-t_{i+1})\,
      e^{-i\omega_i(t_i-t_{i+1})}
      \left\langle \left\langle \hat  I_i(t_1) \hat I_j(t_2)
      \hat I_k(t_3) \hat I_l(t_4) \right\rangle\right\rangle,
      \label{S4}
\end{equation}
whereas the reducible ones can be expressed in terms of the noise
spectral density $S_{ij}^{(2)}(\omega)$ (cf. eq.~(\ref{S2})).
Introducing a damping factor and performing the time integration,
we obtain:
\begin{eqnarray}
      &&\langle\langle \hat q^4 \rangle\rangle = 24\sum_{ijkl}
      \frac{\alpha_i \alpha_j \alpha_k \alpha_l}{(2L)^4}\times
      \nonumber\\
      &&
      \times\Biggl\{
      \int \frac{d\omega_1 d\omega_2 d\omega_3}{(2\pi)^3}
      \frac{S_{ijkl}^{(4)}(\omega_1, \omega_2,
      \omega_3)}{(\omega_1-\Omega-i\eta)((\omega_2-2\Omega)^2+4\eta^2)
      (\omega_3-\Omega+i\eta)}
      \nonumber\\
      &&+ \int \frac{d\omega_1 d\omega_2}{(2\pi)^2} \,
      \frac{S_{ij}^{(2)}(\omega_1) S_{kl}^{(2)}(\omega_2)}{ (\omega_1\!
      +\omega_2\! - 2\Omega)^2 +4\eta^2} \, \Biggl(
      \frac1{(\omega_2\!-\Omega)^2 +\eta^2}
      \nonumber\\
      &&+\frac1{(\omega_1\!-\Omega + i\eta) (
      \omega_2\!-\Omega -i\eta)}\Biggr) \Biggr\}
      - 3\langle \hat q^2\rangle^2.
      \label{q4}
\end{eqnarray}

Finally, we note that the output of a measurement of the power
spectrum of current fluctuations performed with the same LC
circuit at several points contains not only cross-covariances
($S_{ij}(\pm\Omega)$ for $i\neq j$), but also auto covariances
($S_{ii}(\pm\Omega)$ at the same point). To measure only
cross-covariances, one must use two independent LC circuits with
equal resonant frequencies coupled to the conductor at only one
point. The results presented above can be extended to the case of
several detectors.

In particular, for two LC circuits coupled to the conductor at
points $x$ and $y$, respectively, it can be shown that the charge
correlator $\langle \hat q_x(t_1) \hat q_y(t_1) \rangle$
calculated to second order of perturbation theory is proportional
to the second-order cross-covariance of current fluctuations:
\begin{equation}
      \langle \hat q_x \hat q_y \rangle = \frac{\alpha_x
      \alpha_y}{(2L)^2}\, \frac1{\eta} \Bigl\{
      (N+1)\bigl(
      S_{xy}^{(2)}(\Omega)+ S_{yx}^{(2)}(\Omega) \bigr)
      - N \bigl( S_{xy}^{(2)}(-\Omega)+
      S_{yx}^{(2)}(-\Omega) \bigr) \Bigr\}.
\end{equation}

\section{MEASUREMENT OF UNEQUAL-TIME CURRENT FLUCTUATIONS}

Consider an experiment where current correlators are to be
measured at certain points in time. In a thought experiment of
this kind, we can use single-electron ammeter, whose readout is
proportional to the exact instantaneous value of current. As a
classical model of the ammeter, we can consider a damped harmonic
oscillator driven by an external force proportional to current:
\begin{equation}
      \ddot{x}(t) +\eta \dot{x}(t) + \Omega^2 x(t) =
      \alpha\, I(t),
\end{equation}
where $x(t)$ is the ammeter readout, $\Omega$ is the resonant
frequency, $\eta$ is the damping factor, and $\alpha$ is the
coupling constant. The solution to this equation can be
represented in terms of the response function $x(t) = \int
\chi(t-t^\prime) I(t^\prime) dt^\prime$:
\begin{equation}
      \chi(t) = \int \frac{d\omega}{2\pi} \chi(\omega) e^{-i\omega
      t}
      = \int \frac{d\omega}{2\pi} \frac{-\alpha\, e^{-i\omega t}}{
      \omega^2-\Omega^2 + i\eta \omega}.
      \label{resp}
\end{equation}

Since the oscillator motion is described by a linear equation, we
can use the classical dynamical susceptibility $\chi(\omega)$ of
the damped oscillator to determine temperature Green’s functions
by invoking the fluctuation– dissipation theorem~\cite{weiss}.
Accordingly, we first express the correlation function $\langle
x(t_1) ... x(t_n)\rangle$ of detector outputs in terms of Green’s
functions for the free oscillator and then replace these functions
with those for the damped oscillator.

Thus, we assume that the detector output is the ammeter readout
$\hat x(t)$, and the ammeter is described by the Hamiltonian:
\begin{equation}
      \hat H_\mathrm{det} = \frac{\hat p^2}{2m} + \frac{m\Omega^2 \hat
      x^2}{2} - \alpha m\, \hat x \hat I(t),
\end{equation}
where the “mass” $m$ and the momentum operator $\hat p$ play the
roles of moment of inertia and angular momentum, respectively.
Unlike LC circuit Hamiltonian~(\ref{hLC}) it is linear in the
coordinate rather than momentum of the oscillator.

In the coordinate representation, the amplitude of the transition
from the state $(x_1,t_1)$ to the state $(x_2,t_2)$ driven by the
external current $\hat I(t)$ is expressed as
follows~\cite{fenver}:
\begin{eqnarray}
      &&\hat S(x_1,t_1;x_2,t_2) = S_0(x_1,x_2,T)\,
      {\cal T} \exp\Biggl(\frac{im\Omega}{\hbar\sin\Omega T}
      \Biggl\{ \frac{\alpha x_1}{\Omega} \int\limits_{t_1}^{t_2}
      \hat I(t) \sin\Omega(t_2-t) dt
      \label{Sa}\\
      &&+ \frac{\alpha x_2}{\Omega} \int\limits_{t_1}^{t_2}
      \hat I(t) \sin\Omega(t-t_1) dt - \frac{\alpha^2}{\Omega^2}
      \int\limits_{t_1}^{t_2} dt \int\limits_{t_1}^t ds\,
      \hat I(t) \hat I(s)\, \sin\Omega(t_2-t)
      \sin \Omega(s-t_1)\Biggr\}\Biggr),
      \nonumber
\end{eqnarray}
where $S_0(x_1,x_2,T)$ is the transition amplitude for a free
oscillator given by~(\ref{S0}).

As shown above (see eq.~(\ref{D})), in the case of an ideal
measurement of the coordinate $\hat x(t_i)$ at each point in time,
the correlation function $\langle x(t_1)...x(t_n)\rangle$ is
calculated as the average over the intermediate coordinates of
evolution operators of the form: $$\int x_i dx_i\, \hat
S_{i,i+1}^\dagger \hat S_{i-1,i}^\dagger ... \hat S_{i-1,i}\hat
S_{i,i+1},$$ where $\hat S_{ij} = \hat S(x_i,t_i; x_j,t_j)$.
However, it is clear from expression~(\ref{Sa}) for the evolution
operator that such an average is not well defined. Indeed, as a
function of $x_i$ this product of evolution operators contains the
factor:
\begin{eqnarray}
      &&{\cal T}_\pm \exp \Biggl( \frac{i\alpha m x_i}{\hbar}
      \Biggl\{ \int_{t_{i-1}}^{t_i} \bigl(
      \hat I_{\scriptscriptstyle +}(t) - \hat
      I_{\scriptscriptstyle -}(t) \bigr)
      \frac{\sin\Omega(t-t_{i-1})}{\sin\Omega(t_i-t_{i-1})}\,
      dt
      \nonumber\\
      &&\qquad\qquad
      + \int_{t_i}^{t_{i+1}} \bigl( \hat I_{\scriptscriptstyle
      +}(t) - \hat I_{\scriptscriptstyle -}(t) \bigr)
      \frac{\sin\Omega(t_{i+1}-t)}{\sin\Omega(t_{i+1}-t_i)}\,
      dt \Biggr\}\Biggr),
\end{eqnarray}
where ${\cal T}_\pm$ denotes the time-ordering operators
introduced to order the current operators $\hat
I_{\scriptscriptstyle +}(t)$ and $\hat I_{\scriptscriptstyle
-}(t)$ forward and backwards in time, respectively. The current
operator under the time-ordering symbol can be treated as ordinary
functions of time. This demonstrates that a weighted average of
this form of the coordinate $x_i$ is a divergent function of $\hat
I_{\scriptscriptstyle \pm}(t)$.

One possible explanation of this divergency is that the oscillator
with exactly known coordinate $x_i$ has infinite energy.
Accordingly, the probability that the oscillator will have a certain
coordinate $x_{i+1}$ at a subsequent point in time $t$ is
independent of $x_{i+1}$ $$P(x_{i+1}|x_i) = \frac{m\Omega}{2\pi
\hbar \sin\Omega t},$$ (e.g. see~\cite{fenver}).

This unphysical result can be avoided by several means. First, we
can assume that the coordinate is measured with limited accuracy
$|x-x_i|\sim\Delta$. Then, the postmeasurement state of the
oscillator is described by a density matrix localized within the
measurement error, such as $\rho_{x_i}(x,y) \sim
\exp(-[(x-x_i)^2+(y-x_i)^2]/4\Delta^2)$. Second, as noted above, the
measurement maps the detector density matrix to a certain “standby”
state.

Hereinafter, we consider the latter scenario and assume that the
measurement on $\hat x(t_i)$ brings the detector into an equilibrium
state at temperature $\Theta$. Then, the correlation function of
detector outputs is given by formal expression~(\ref{xxw}).
Integrating the evolution operators over the initial and final
coordinates, we obtain:
\begin{equation}
      \langle x(t_1) ... x(t_n) \rangle =\left\langle
      {\cal T}_K F_K[\hat I]\prod_{i=1}^n \Bigl( \frac{i\alpha}{\hbar}
      \int\limits_{K_{i-1}^i} C_K(t_i-t) \hat I(t)\, dt \Bigr)
      \right\rangle
      \label{xx}
\end{equation}
where the Keldysh Green’s function is defined as: $C_K(t-s)=\langle
{\cal T}_K \{ \hat x(t) \hat x(s) \} \rangle$;
$K_i^j=[t_i+i0\rightarrow t_j \rightarrow t_i-i0]$ - is the Keldysh
contour. The influence functional $F_K[\hat I]$:
\begin{equation}
      F_K[\hat I] = \exp \left[ - \frac{\alpha^2}{2\hbar^2}
      \mathop{\int\int}_K C_K(t-s) \hat I(t) \hat I(s) \, dt ds
      \right],
\end{equation}
is defined on the Keldysh contour $K_{-\infty}^{\infty}$.

The Keldysh Green’s function $C_K(t)$ can be expressed in terms of
the causal Green’s function $C(t>0)=\langle \hat x(t) \hat x(0)
\rangle$, which can in turn be expressed in terms of the dynamical
susceptibility given by $\chi(\omega)$~(\ref{resp}), by invoking the
fluctuation–dissipation theorem:
\begin{equation}
      C(t) = \frac{\hbar}{\pi} \int d\omega\, e^{-i\omega t} \,
      \frac{\mbox{Im} \chi(\omega)}{1-e^{-\hbar\omega/\Theta}}.
      \label{fdt}
\end{equation}

In what follows, Green’s function $C(t)$ is conveniently represented
as the complex variable: $C(t) = S(t)+i A(t)$, with real and
imaginary parts corresponding to the symmetrized and antisymmetrized
coordinate autocorrelation functions: $S(t) = \frac12\langle \hat
x(t) \hat x(0) + \hat x(0) \hat x(t) \rangle$, and $A(t) =
\frac1{2i} \langle \hat x(t) \hat x(0) - \hat x(0) \hat x(t)
\rangle$. Using the relation $1/(1-e^{-x})= \frac12 + \frac12
\coth(x/2)$, we obtain:
\begin{equation}
      S(t) = \frac{\hbar}{2\pi} \int d\omega\, \cos(\omega t) \,
      \mbox{Im} \chi(\omega)\, \coth\Bigl( \frac{\hbar \omega}{2
      \Theta} \Bigr),
      \label{xxs}
\end{equation}
\begin{equation}
      A(t) = -\frac{\hbar}{2\pi} \int d\omega \,
      \sin(\omega t)\, \mbox{Im}\chi(\omega).
      \label{xxa}
\end{equation}

\subsection{Measurement of Second-Order Current
Correlators}

Here, we derive an expression for the correlation function $\langle
x(t_1) x(t_2) \rangle$ in a perturbation theory in the coupling
constant $\alpha\ll 1$ thus neglecting the back-action of the
detector on conductor dynamics (i.e., by setting $F_K[\hat I]=1$).
The perturbative calculation of $\langle x(t_1) x(t_2) \rangle$ was
first carried out in Ref.~\cite{lesovik2}. Rewriting Eq.~(\ref{xx})
in explicitly time-ordered form, we find~\cite{lesovik2}
\begin{eqnarray}
      &&\langle x(t_1) x(t_2) \rangle =
      \frac{4\alpha^2}{\hbar^2}
      \int\limits_{t_1}^{t_2} dt_2^\prime
      \int\limits_{t_0}^{t_1} d t_1^\prime\,
      \Bigl\{ A(t_2-t_2^\prime) A(t_1-t_1^\prime)
      \bigl\langle \hat I(t_2^\prime) \hat I(t_1^\prime)
      \bigr\rangle^S
      \nonumber\\
      &&\qquad\qquad\qquad\qquad
      - i A(t_2-t_2^\prime) S(t_1-t_1^\prime)
      \bigl\langle \hat I(t_2^\prime) \hat I(t_1^\prime)
      \bigr\rangle^A \Bigr\},
      \label{x2}
\end{eqnarray}
where $\langle \hat I(t_1) \hat I(t_2) \rangle^S$ and $\langle \hat
I(t_1) \hat I(t_2) \rangle^A$ are the symmetrized and
antisymmetrized current correlators,
\begin{eqnarray}
      &&\langle \hat I(t_1) \hat I(t_2) \rangle^S= \frac12
      \langle \hat I(t_1) \hat I(t_2) + \hat I(t_2)
      \hat I(t_1) \rangle,
      \\
      &&\langle \hat I(t_1) \hat I(t_2) \rangle^A = \frac12 \langle
      \hat I(t_1) \hat I(t_2) - \hat I(t_2) \hat I(t_1) \rangle.
\end{eqnarray}

Since we focus on the measurement of current correlators at fixed
times, $t_1$ and $t_2$, the functions $A(t)$ and $S(t)$ in
Eq.~(\ref{x2}) must decay over time intervals much shorter than the
time scale $\tau_e$, of the evolution of the electron system.
Otherwise, the correlator $\langle x(t_1)x(t_2)\rangle$ would be
proportional to time-averaged current correlators. Calculating the
integral over frequency in Eq.~(\ref{xxa}), we find that $A(t)$ is a
decaying oscillating function independent of temperature:
\begin{equation}
      A(t) = - \frac{\hbar}{2\omega_\eta}\, \exp\Bigl( -\frac{\eta
      |t|}2 \Bigr)
      \sin (\omega_\eta  t),
\end{equation}
where $\omega_\eta = \sqrt{\Omega^2-\eta^2/4}$ - is the renormalized
oscillator frequency.

The symmetrized correlation function $S(t)$ is the sum of two
functions, $S(t)=S_1(t)+S_2(t)$, having different long-time
behavior. The decay of $S_1(t)$ is completely determined by the
oscillator parameters:
\begin{equation}
      S_1(t) = \frac{\hbar}{4\omega_\eta}\,e^{-\eta t/2}\Bigl\{
      e^{i\omega_\eta t} \coth\frac{\hbar (\omega_\eta +i
      \frac{\eta}{2})}{2\Theta} + C.c. \Bigr\},
\end{equation}
whereas the decay of the other function depends on the oscillator
temperature:
\begin{equation}
      S_2(t) = - 2\eta \Theta \sum_{n=1}^\infty \frac{\nu_n
      e^{-\nu_n t}}{(\nu_n^2+\Omega^2)^2-\eta^2\nu_n^2},
      \label{xxs2}
\end{equation}
where $\nu_n = 2\pi n \Theta/\hbar$ is a Matsubara frequency.

Since the oscillator frequency is renormalized with $\Omega$ held
constant, the functions $A(t)$ and $S_1(t)$ cannot decay faster than
$e^{-\Omega t}$ (at $\eta=2\Omega$) as $t\rightarrow \infty$. In the
strong-damping limit ($\eta\gg \Omega$), the damping factor is
smaller, and both $A(t)$ and $S_1(t)$ decay approximately as $\sim
e^{-\Omega (\Omega/\eta)t}$. Therefore, measurements of unequal-time
correlators should be performed with oscillator parameters chosen so
that $\eta\sim2\Omega \gg \max \{ \tau_e^{-1}, |t_2-t_1|^{-1}\}$.

Since the decay of $S_2(t)$ depends on the detector temperature, it
is necessary that $\nu_1\gg \tau_e^{-1}$, which is equivalent to the
requirement that $\Theta\gg \max\{\hbar\tau_e^{-1},
\hbar|t_2-t_1|^{-1}\}$. For a ballistic conductor, the time scale of
current fluctuations is $\tau_e\sim \hbar/eV$, where $eV$ is the
voltage across the conductor; therefore, the condition $\Theta\gg
\hbar\tau_e^{-1}$ corresponds to $\Theta\gg eV$. This means that
weakly nonequilibrium current fluctuations will be measured when the
detector and the conductor have comparable temperatures.

In this approximation, the current operators in~(\ref{x2}) can be
factored out from the integral, and we can set:
\begin{eqnarray}
      &&\int_{t_1}^{t_2} A(t_2-t)\,
      \hat I(t)\, dt \approx -\frac{\hbar}{2\Omega^2}\, \hat I(t_2).
      \label{Ai}
      \\
      &&\int_{t_1}^{t_2} S(t_2-t)\, \hat I(t)\, dt \approx
      \frac{\eta\Theta}{\Omega^4}\, \hat I(t_2).
      \label{Si}
\end{eqnarray}
As a result, we find that the quantity measured at high temperatures
is the following combination of the symmetrized and antisymmetrized
current correlators:
\begin{equation}
      \langle x(t_1)x(t_2) \rangle=\frac{\alpha^2}{\Omega^4}
      \Bigl\{ \bigl\langle \hat I(t_2) \hat I(t_1) \bigr\rangle^S
      +i \Bigl( \frac{2\eta\Theta}{\hbar\Omega^2} \Bigr)
      \bigl\langle \hat I(t_2) \hat I(t_1) \bigr\rangle^A \Bigr\}.
      \label{x2high}
\end{equation}
At $\eta\sim 2\Omega$ the relative contribution of either correlator
depends on the ratio $4\Theta/\hbar\Omega$, which can have any value
in this regime.

The high temperature result~(\ref{x2high}) in the limit
$\omega_\eta\ll \eta$ was obtained in Ref.~\cite{lesovik2}. Another
interesting experiment is measurement of quantum current
fluctuations in the low-temperature regime ($\Theta\ll
\hbar\tau_e^{-1}$). In this regime, measurement of highly
nonequilibrium excess fluctuations becomes particularly feasible. In
contrast to the high-temperature regime, the long-time decay of
$S(t)$ follows a power low,
\begin{equation}
      S(t\gg \Omega^{-1}) = -\Bigl( \frac{\hbar \eta}{ \pi \Omega^4}
      \Bigr)\, \frac1{t^2}.
\end{equation}

Again, requiring that $\eta\sim\Omega \gg \max\{\tau_e^{-1},
|t_2-t_1|^{-1}\}$, we can assume,
\begin{equation}
      \int\limits_{t_1}^{t_2} S(t_2\!-\!t) \hat I(t)\, dt
      \approx \hat{I}(t_2)\!\!\int\limits_{0}^{t_2-t_1} dt\, S(t)
      \approx \frac{\eta}{\Omega^4} \Biggl( \Theta +
      \frac{\hbar}{\pi (t_2\!-\!t_1)} \Biggr) \hat{I}(t_2),
\end{equation}
where in the last equation we have used the relation
\begin{equation}
      \int_0^\infty S(t)dt = \eta\Theta/\Omega^4,
\end{equation}
which holds at any temperature. The resulting expression for
$\langle x(t_1) x(t_2) \rangle$ at low temperature of the detector
takes the form,
\begin{equation}
      \langle x(t_1) x(t_2) \rangle = \frac{\alpha^2}{\Omega^4}
      \Bigl\{ \bigl\langle \hat I(t_2) \hat I(t_1) \rangle^S +
      i\frac{2\eta}{\hbar\Omega^2} \Bigl( \Theta + \frac{\hbar}{
      \pi(t_1-t_0)} \Bigr) \bigl\langle \hat I(t_2) \hat
      I(t_1) \bigr\rangle^A \Bigr\}
      \label{x2low}
\end{equation}

Comparing the last expression with the high-temperature
result~(\ref{x2high}) we conclude that in the low-temperature limit
the contribution of the antisymmetrized current correlator to the
$\langle x(t_1) x(t_2) \rangle$ is proportional to the effective
temperature,
\begin{equation}
      \Theta^\prime = \Theta + \frac{\hbar}{\pi\Delta t},
\end{equation}
where $\Delta t$ is the time separation between subsequent
measurement times.

\subsection{Measurement of Higher Order Correlators}

Perturbative treatment of Eq.~(\ref{xx}) makes it possible to
express a coordinate correlation function of any order in terms of
current correlators. First, consider measurement of third-order
correlators. Time-ordering the current operators along the Keldysh
contour, we obtain
\begin{eqnarray}
      &&\langle x(t_1) x(t_2) x(t_3) \rangle =
      -\Bigl(\frac{2\alpha}{\hbar} \Bigr)^3
      \int_{t_2}^{t_3} dt_3^\prime
      \int_{t_1}^{t_2} dt_2^\prime
      \int_{t_0}^{t_1} dt_1^\prime \times
      \nonumber\\
      &&\times
      \Bigl\{
      A(t_3\!-t_3^\prime) A(t_2\!-t_2^\prime) A(t_1\!-t_1^\prime) \,
      \bigl\langle
      \hat I(t_3^\prime) \hat I(t_2^\prime) \hat I(t_1^\prime)
      \bigr\rangle^S
      \nonumber\\
      &&
      -i A(t_3\!-t_3^\prime) A(t_2\!-t_2^\prime) S(t_1\!-t_1^\prime)\,
      \bigl\langle
      \hat I(t_3^\prime)\hat I(t_2^\prime) \hat I(t_1^\prime)
      \bigr\rangle^{SA_1}
      \nonumber\\
      &&-i A(t_3\!-t_3^\prime) S(t_2\!-t_2^\prime) A(t_1\!-t_1^\prime) \,
      \bigl\langle
      \hat I(t_3^\prime) \hat I(t_2^\prime) \hat I(t_1^\prime)
      \bigr\rangle^{SA_2}
      \nonumber\\
      &&
      + A(t_3\!-t_3^\prime) S(t_2\!-t_2^\prime) S(t_1\!-t_1^\prime) \,
      \bigl\langle
      \hat I(t_3^\prime) \hat I(t_2^\prime) \hat I(t_1^\prime)
      \bigr\rangle^{A} \Bigr\}.
      \label{x3}
\end{eqnarray}

It is clear that four different third-order current correlators are
measured in this case. Denoting by $\{A,B\}$ and $[A,B]$ the
anticommutator and commutator of operators, respectively, we
represent two of the four correlators as averages of symmetrized and
antisymmetrized combinations of current operators taken at different
times,
\begin{eqnarray}
      &&\bigl\langle \hat I_3 \hat I_2 \hat I_1
      \bigr\rangle^S = \frac14 \left\langle \bigl\{ \bigl\{ \hat
      I_3, \hat I_2 \bigr\}, \hat I_1 \bigr\}
      \right\rangle,
      \label{s3meas}
      \\
      &&\bigl\langle \hat I_3 \hat I_2 \hat I_1
      \bigr\rangle^{A} = \frac14 \left\langle \bigl[\, \bigl[ \hat
      I_3, \hat I_2 \,\bigr], \hat I_1 \,\bigr]
      \right\rangle,
      \label{AA}
\end{eqnarray}
and the other two as averages of mixed combinations of symmetrized
and antisymmetrized current operators,
\begin{eqnarray}
      &&\bigl\langle \hat I_3 \hat I_2 \hat I_1
      \bigr\rangle^{SA_1} = \frac14 \left\langle \bigl[ \bigl\{ \hat
      I_3, \hat I_2 \bigr\}, \hat I_1 \bigr]
      \right\rangle.\qquad
      \label{SA}
      \\
      &&\bigl\langle \hat I_3 \hat I_2 \hat I_1
      \bigr\rangle^{SA_2} = \frac14 \left\langle \bigl\{ \bigl[ \hat
      I_3, \hat I_2 \bigr], \hat I_1 \bigr\}
      \right\rangle.\qquad
      \label{AS}
\end{eqnarray}

Note that these four Hermitian combinations of current operators are
not the only possible ones. For example, following a standard
prescription, we might assume that the completely symmetrized
combination is measurable:
\begin{equation}
      \bigl\langle \hat I_1 \hat I_2 \hat I_3 \bigr\rangle^{S^\prime} =
      \frac16 \Bigl\langle \hat I_1\! \bigl\{ \hat I_2,\hat I_3
      \bigr\}\! + \hat I_2\! \bigl\{ \hat I_3, \hat I_1\bigr\}\! + \hat
      I_3\! \bigl\{ \hat I_2,\hat I_1 \bigr\}  \Bigr\rangle.
      \label{s3full}
\end{equation}
However, the analysis above shows that this is not true. The formal
reason is that the current operators used to describe detector
dynamics are Keldysh time-ordered. Since $t_1<t_2<t_3$, the
combinations $\langle \hat I(t_2) \hat I(t_1) \hat I(t_3) \rangle$
and $\langle \hat I(t_3) \hat I(t_1) \hat I(t_2) \rangle$, which are
contained in~(\ref{s3full}) and not contained in~(\ref{s3meas}), can
not appear as a result of such time-ordering. Indeed, the operator
$\hat I(t_1)$, to be measured first will be on the right or left of
other current operators, depending on whether time- or
antitime-ordering is performed, respectively. In other words, the
absence of $\langle \hat I(t_2) \hat I(t_1) \hat I(t_3) \rangle$ and
$\langle \hat I(t_3) \hat I(t_1) \hat I(t_2) \rangle$ in
correlator~(\ref{s3meas}) is dictated by causality: measurements
performed at later points in time $t_2$ and $t_3$ cannot affect the
outcome of the earlier measurement at $t_1$.

This argumentation suggests that a linear detector can be used to
measure only correlators of the form:
\begin{equation}
      \langle \hat x(t_1)...x(t_n)\rangle \sim
      \Bigl\langle \bigl\{\bigl\{\hat I(t_n),\hat I(t_{n-1})
      \bigr\},...\bigr\},\hat I(t_1)\bigr\} \Bigr\rangle,
      \label{xn}
\end{equation}
where $t_1\leq ... \leq t_n$, as well as the remaining $2^{n-1}-1$
correlators obtained by replacing any number of anticommutators with
commutators.

Note that, depending on the measurement regime, the contributions of
some of these $2^{n-1}$ correlators to $\langle
x(t_1)...x(t_n)\rangle$ may be equal. In particular, consider the
measurement of a third-order correlator,
\begin{eqnarray}
      &&\langle x(t_1) x(t_2) x(t_3) \rangle =
      \frac{\alpha^3}{\Omega^6} \Bigl\{ \bigl\langle \hat
      I(t_3) \hat I(t_2) \hat I(t_1) \bigr\rangle^S + i \Bigl(
      \frac{2\eta\Theta_1^\prime}{\hbar\Omega^2} \Bigr) \bigl\langle \hat I(t_3)
      \hat I(t_2) \hat I(t_1) \bigr\rangle^{SA_1}
      \nonumber\\
      &&+ i \Bigl(
      \frac{2\eta\Theta_2^\prime}{\hbar\Omega^2} \Bigr) \bigl\langle \hat I(t_3)
      \hat I(t_2) \hat I(t_1) \bigr\rangle^{SA_2}
      + \Bigl(\frac{2\eta\Theta_2^\prime}{\hbar\Omega^2} \Bigr) \Bigl(
      \frac{2\eta\Theta_1^\prime}{\hbar\Omega^2} \Bigr) \bigl\langle \hat I(t_3)
      \hat I(t_2) \hat I(t_1) \bigr\rangle^{A} \Bigr\},
\end{eqnarray}
where $\Theta_1^\prime = \Theta+\frac{\hbar}{\pi(t_1-t_0)}$ and
$\Theta_2^\prime = \Theta+\frac{\hbar}{\pi(t_2-t_1)}$ are two
effective temperatures. In the high-temperature regime one has
$\Theta_1^\prime\approx\Theta_2^\prime\approx\Theta$ and hence,
\begin{eqnarray}
      &&\langle x(t_1) x(t_2) x(t_3) \rangle =
      \frac{\alpha^3}{\Omega^6} \Bigl\{
      \bigl\langle
      \hat I(t_3) \hat I(t_2) \hat I(t_1)
      \bigr\rangle^S
      \nonumber\\
      &&\qquad
      + i \Bigl( \frac{2\eta\Theta}{\hbar\Omega^2} \Bigr)
      \bigl\langle
      \hat I(t_3) \hat I(t_2) \hat I(t_1)
      \bigr\rangle^{SA_3}
      + \Bigl( \frac{2\eta\Theta}{\hbar\Omega^2} \Bigr)^2
      \bigl\langle
      \hat I(t_3) \hat I(t_2) \hat I(t_1)
      \bigr\rangle^{A} \Bigr\},
\end{eqnarray}
where the correlator $\langle \hat I_3 \hat I_2 \hat I_1
\rangle^{SA_3}$ is a combination of $\langle \hat I_3 \hat I_2 \hat
I_1 \rangle^{SA_1}$ and $\langle \hat I_3 \hat I_2 \hat
I_1\rangle^{SA_2}$ having the form:
\begin{equation}
      \bigl\langle \hat I_3 \hat I_2 \hat I_1 \bigr
      \rangle^{SA_3} = \frac12\,\Bigl( \bigl\langle \hat I_3
      \hat I_2 \hat I_1 \bigr\rangle - \bigl\langle \hat I_1
      \hat I_2 \hat I_3 \bigr\rangle \Bigr).
\end{equation}

\section{CONCLUSIONS}

We have theoretically analyzed measurements of higher-order current
correlators with a quantum detector coupled to the conductor. A
damped harmonic oscillator with coordinate $\hat x(t)$ measured
directly is considered as a model of a detector whose classical
dynamics are described by an equation of motion linear in $x$.

Two essentially different types of measurement have been considered:
measurement of the power spectrum of current fluctuations and
measurement of unequal-time current correlators at several points in
time.

In the former measurement, a weakly damped resonant LC circuit
inductively coupled to the conductor is used as a detector. In
measurements of this type, the detector interacts with the conductor
over a considerable time interval, and the measured observable $\hat
q$ is the charge stored in the capacitor of the LC circuit.
Measurement of moments $\langle \hat q^n\rangle$ of the stored
charge provides information about $n$th-order power spectrum of
current fluctuations. A complete statistical description of $\hat q$
is given by Eq.~(\ref{chiq}) in terms of integrals of Keldysh
time-ordered current operators.

This equation is used to develop a perturbation theory in the
coupling strength. By neglecting the backaction of the detector on
the measured system, expression (\ref{q2}) is obtained for the
second-order irreducible charge correlator $\langle\langle \hat
q^2\rangle\rangle$, in the case when the detector is coupled to the
conductor at several points. At zero detector temperature,
$\langle\langle \hat q^2\rangle\rangle$ is proportional to the
positive frequency Fourier transform of the second order current
correlator symmetrized with respect to coordinates. At a finite
detector temperature, negative-frequency current correlator
symmetrized with respect to coordinates also contributes to the
stored charge.

Measurements of unequal-time current correlators are analyzed in the
strong-damping limit, when the damping factor comparable to the
oscillator frequency. The measured quantity is the correlation
function $\langle x(t_1)...x(t_n) \rangle$ of the oscillator
coordinate at subsequent points in time. The oscillator is assumed
to relax to equilibrium after each measurement. This type of
measurement is different from von Neumann’s projective measurements
at subsequent time moments. It is found that the correlation
function $\langle x(t_1)...x(t_n) \rangle$ may be difficult to
determine because an electron in a sharply localized state has
infinite energy.

In the general case, the outcome of time resolved measurements of
the detector state is described by Eq.~(\ref{xx}). Perturbation
theory is used to derive expressions~(\ref{x2}) and~(\ref{x3}) for
second- and third-order coordinate correlators, respectively. When
second-order correlators are measured, the detector output combines
contributions of current correlators symmetrized and antisymmetrized
with respect to time. In the high-temperature regime ($\Theta \gg
eV$) the contribution to $\langle x(t_1)x(t_2)\rangle$ due to the
antisymmetrized current correlator is proportional to the detector
temperature $\Theta$, see Eq.~(\ref{x2high}). At low temperatures,
($\Theta\ll eV$) one has the same result for $\langle
x(t_1)x(t_2)\rangle$ as in the high-temperature regime with an
effective detector temperature $\Theta^\prime=\Theta
+\hbar/(\pi\Delta t)$, where $\Delta t$ is a time separation between
subsequent measurements, see Eq.~(\ref{x2low}).

When higher order coordinate correlators are measured, a larger
number of differently time-ordered current correlators contribute to
the measurement outcome. The current operators in such correlators
are Keldysh time-ordered in accordance with the causality
requirement: a future current measurement cannot affect the outcome
of a past measurement.

We thank G. Blatter for stimulating discussions. This work was
supported by the Russian Foundation for Basic Research, project no.
06-02-17086-a, by the Russian Academy of Sciences under the program
“Quantum Macrophysics,” and by the Swiss National Foundation.

\end{document}